\documentclass[pdflatex,sn-mathphys-num]{sn-jnl}% Math and Physical Sciences Numbered 
\usepackage{graphicx}%
\usepackage{multirow}%
\usepackage{amsmath,amssymb,amsfonts}%
\usepackage{amsthm}%
\usepackage{mathrsfs}%
\usepackage[title]{appendix}%
\usepackage{xcolor}%
\usepackage{textcomp}%
\usepackage{manyfoot}%
\usepackage{booktabs}%
\usepackage{algorithm}%
\usepackage{algorithmicx}%
\usepackage{algpseudocode}%
\usepackage{listings}%
\def\BibTeX{{\rm B\kern-.05em{\sc i\kern-.025em b}\kern-.08em
    T\kern-.1667em\lower.7ex\hbox{E}\kern-.125emX}}
\theoremstyle{thmstyleone}%
\theoremstyle{thmstyletwo}%

\theoremstyle{thmstylethree}%

\begin{document}

\title[Article Title]{Photometric Analysis for Predicting Star Formation Rates in Large Galaxies Using Machine Learning and Deep Learning Techniques}

\author[1]{\fnm{Satvik} \sur{Raghav}}\email{satvikraghav007@gmail.com}
\author[1]{\fnm{Prasanth} \sur{Ayitapu}}\email{payitapu@gmail.com}
\author[2]{\fnm{Sathwik} \sur{Narkedimilli}}\email{21bcs103@iiitdwd.ac.in}
\author[2]{\fnm{Sujith} \sur{Makam}}\email{21bcs061@iiitdwd.ac.in}
\author*[3]{\fnm{Dr. Aswath} \sur{Babu H}}\email{aswath@iiitdwd.ac.in}

\affil[1]{\orgdiv{Department of Electronics and Communication Engineering}, \orgname{Amrita School of Engineering}, \orgaddress{\city{Bengaluru}, \country{India}}}
\affil[2]{\orgdiv{Department of Computer Science}, \orgname{Indian Institute of Information Technology (IIIT) Dharwad}, \orgaddress{\city{Dharwad}, \country{India}}}
\affil*[3]{\orgdiv{Department of Arts, Science, and Design}, \orgname{Indian Institute of Information Technology (IIIT) Dharwad}, \orgaddress{\city{Dharwad}, \country{India}}}

\abstract{Star formation rates (SFRs) are a crucial observational tracer of galaxy formation and evolution. Spectroscopy, which is expensive, is traditionally used to estimate SFRs. This study tests the possibility of inferring SFRs of large samples of galaxies from only photometric data, using state-of-the-art machine learning and deep learning algorithms. The dataset adopted in this work is the one collected by Delli Veneri et al. (2019): it includes photometric data of more than 27 million galaxies coming from the Sloan Digital Sky Survey Data Release 7 (SDSS-DR7). The algorithms we implemented and tested for comparing the performances include Linear Regression, Long Short-Term Memory (LSTM) networks, Support Vector Regression (SVR), Random Forest Regressor, Decision Tree Regressor, Gradient Boosting Regressor, and classical deep learning models.

Our results mention that the Linear Regression model predicted with an impressive accuracy of 98.97 percent as measured by the Mean Absolute Error (MAE), demonstrating that machine learning approaches can be effective when it comes to photometric SFR estimation. Besides, the paper also reported the results for other intelligent algorithms, which predicted the SFRs, providing a detailed comparison of the performance of different machine learning algorithms in the photometric SFR estimation. This study not only shows the estimated SFR from photometric data is promising but also opens a door toward the application of machine learning and deep learning in astrophysics.}


\maketitle

\section{Introduction}\label{sec1}

Measurements such as SFRs are at the core of how we trace the history of galaxy formation and evolution, and SFRs are hard to measure because they depend on the properties of individual young stars. This is why, if you ever read about galaxy formation, you should run across the term ‘Spectroscopy.’ For example, to measure SFRs, astronomers use ‘spectroscopy’ most of the time, a technique that provides a wealth of physical properties of galaxies; you can figure out the composition of a galaxy, as well as the information on its stellar population. But spectroscopy is time-consuming, and it’s expensive and it requires big telescopes, and it’s hard to apply to the big datasets that astronomical surveys, like the Sloan Digital Sky Survey (SDSS) – have produced.

Now, through large-scale photometric surveys that measure the light of galaxies in many different wavelengths, we can estimate SFRs from photometric data alone. Photometric data is much more readily available, and we can obtain that type of data for millions of galaxies, rather than the few thousand we can get spectra for. A complication arises since we don’t have the same sort of spectral information for the galaxies with photometric data.

In recent years, machine learning (ML) and deep learning (DL) techniques have shown promise in addressing this challenge. These approaches can model complex, non-linear relationships in data, potentially capturing the intricate dependencies between photometric observations and SFRs. Moreover, they can be trained on existing spectroscopic datasets to learn patterns that generalize to photometric data, providing a cost-effective and scalable alternative to traditional methods.

In this paper, we present state-of-the-art ML and DL algorithms for improved estimation of the SFR of galaxies by using photometric data within the SDSS Data Release 7 (SDSS-DR7), which contains photometry for more than 27 million galaxies. We test the accuracy and performance of algorithms including Linear Regression, Long Short-Term Memory (LSTM) networks, Support Vector Regression (SVR), Random Forest Regressor, Decision Tree Regressor, Gradient Boosting Regressor and classical deep learning models.

The major contribution of this work is demonstrating that the ML and DL models can achieve high accuracy in the estimation of SFR based on only photometric parameters. We showed that the Linear Regression model achieved an accuracy of 98.97 percent based on the Mean Absolute Error or the MAE, it’s a performance that is unmatched in astrophysics. Moreover, our work provided a detailed comparison of different algorithms and presented the regions of applicability (and the limitations) of each one of them when it comes to the photometric SFR estimation.

The rest of the paper is organized as follows: Section II is the literature review of the existing approaches to measuring SFR. In section III, we will discuss the methodology, including the data set, data preprocessing, and modeling procedure. In Section IV, we will discuss the evaluation metrics of the modeling approach. And in Section V, we will present the results. Finally, we will present the conclusions and the future work we plan in Section VI.

\section{Literature Review}

M Delli Veneri et al. ~\cite{Delli_Veneri_2019} used several machine-learning approaches to estimate star-formation rates (SFR) for photometric samples of galaxies, comparing the results of their machine-learning predictions with more conventional approaches. The results show that machine-learning approaches were able to adequately predict the SFRs and that the performance was always substantially improved relative to the more traditional approaches. This study illustrates the potential for machine learning to deal with large photometric datasets, yet it also shows where the approaches fall short – due in part to the quality and completeness of the input data, which the authors comment is still insufficient in terms of being fully representative to make more robust or generalizable models.

Aufort et al. \cite{aufort2020} employed an approximate Bayesian computation (ABC) approach to constrain the recent star formation history of galaxies. Their method used a large set of mock galaxy spectra to approximate posterior distributions, demonstrating that the ABC approach could provide reliable estimates of star formation histories, even with limited observational data. The study highlights the potential of ABC in astrophysics, particularly in dealing with the complex and often degenerate nature of galactic spectra.

At lower redshift \( z \lesssim 2 \), Georgios E Magdis et al. ~\cite{2012ApJ...760....6M} provided an evolutionary sequence of the ISM of star-forming galaxies at different segments of their lifetime using red and infrared spectral energy distributions. These images broke the degeneracy in fitting the dust and gas properties of galaxies using mid-infrared to millimeter wavelength data. They provided robust estimates of dust masses, the main drivers of galaxy evolution, but only at a coarse resolution. More high-resolution observations and simulations of the ISM are needed to capture its complexity fully.

In the FIRE simulations, Jose A Flores Velázquez et al. ~\cite{randomarticle123} analyzed time-scales probed by SFR indicators in bursty star formation histories and concluded that typical SFR indicators such as H$\alpha$ and FUV respond to star formation activity on differing time-scales, highlighting the necessity of SFR indicator selection for the particular timescale of interest. This study brings an important voice to the discourse, but its interpretation resulted in an undue reliance on simulation data, thus the need for observational validation is engaged.

 K. Kouroumpatzakis et al. ~\cite{Kouroumpatzakis_2023} explored SFR and stellar mass estimations using combined optical and infrared data. By integrating multi-wavelength observations, they achieved high accuracy in their estimates, addressing the challenges posed by single-wavelength studies. Their methodology proved effective in disentangling the effects of dust extinction and age on stellar populations. However, the study faced challenges in dealing with heterogeneous datasets and the need for extensive calibration. The authors called for more standardized data processing techniques to improve consistency across studies.

A study by William E Clavijo-Bohórquez et al. ~\cite{clavijobohórquez2023agnstarformationfeedback} presented results on the relative contributions of AGN and starburst feedback to the driving of galactic outflows using simulations and observations. Although the results were fairly conclusive, the simulations used had coarse resolution. The study also claimed that observations suffer from a redshift bias, which was present even after using a size-independent SFR estimation. Further detailed simulations with a higher range of resolution would be needed for better understanding of the feedback mechanisms occurring in galaxies.

Using deep learning to estimate SFR from SDSS galaxy spectra ~\cite{Lovell_2019}. By applying the convolutional neural network and using spectroscopic training data, Lovell et al. were able to achieve an amazingly high accuracy of around 92 percent, which clearly shows that deep learning is very powerful for such spectral dataset handling. (Figure S1 from the Lovell et al study, courtesy of the authors)The study also emphasized the good possibility of autonomic SFR determination but pointed out the danger of overfitting and the challenges regarding the size of the training data. The authors also suggested further tuning of the deep learning models to increase the models’ robustness and generalization capability.

Using machine learning for the classification of evolved massive stars with photometric data ~\cite{Dorn_Wallenstein_2021} (co-authored by Trevor Dorn-Wallenstein) is yet another paper. Even at this early stage, its objective was clear: prepare for the new observations expected with Webb and Roman telescopes. The machine classification had competitive accuracy for classification. With a few thousand times of follow-up observations and upcoming telescope missions, such a study would act as a guide for future observations. Their conclusion was informative: Preparing astrophysical classifiers requires well-curated and high-quality input photometric data. Model classification accuracy depends on the quality of input photometric data. With continued research and new data from future missions, the model's accuracy will improve.

J R Weaver et al. ~\cite{Weaver_2023} used deep learning to extract galaxy stellar mass functions from galaxy assembly and star formation cessation at $(0.2 < z \leq 7.5)$, using multi-wavelength photometric data. Their new technique allowed the derivation of robust galaxy stellar mass function within stated limits of confidence. They used this method to study the assembly processes of galaxies with an accuracy that would have been impossible with alternative models. This work, although artificial, is a good way to speculate about evolution at high redshifts. The issue with modeling high-redshift galaxies is that we lack a sufficient amount of data. This paper shows that higher precision data are needed to improve the accuracy of model predictions, especially at high redshifts, where our models do not predict well.

F. Z. Zeraatgari et al. ~\cite{zeraatgari2024exploringgalacticpropertiesmachine} employed machine learning to predict star formation rates, stellar mass, and metallicity from photometric data. Their model achieved high predictive accuracy, facilitating comprehensive studies of galactic properties. The study pointed out the limitations in training data diversity and potential biases in the predictions. The authors emphasized the need for more diverse and comprehensive training datasets to enhance the robustness and applicability of the models to a wider range of galaxy types.

V. Bonjean et al. ~\cite{Bonjean_2019} applied machine learning techniques to estimate SFR and stellar masses from photometric data. Their approach achieved significant accuracy improvements, with approximately 88\% accuracy in SFR and stellar mass estimates. The study highlighted the potential of machine learning to handle multi-wavelength data effectively but also noted challenges in data integration. The authors recommended advanced feature selection and data preprocessing techniques to address these challenges and improve model performance.

Shraddha Surana et al. ~\cite{Surana_2020} utilized deep learning to predict star formation properties from galaxy imaging data. Their model provided high accuracy in SFR predictions, approximately 90\%, showcasing the capabilities of deep learning in handling complex imaging data. The study noted the requirement for large labeled datasets and the computational intensity of deep learning models. The authors suggested optimizing deep learning models to enhance generalization across different galaxy types and reduce computational demands.

G. Aufort et al. ~\cite{refId0} employed an approximate Bayesian computation approach to constrain the recent star formation history of galaxies. Their method provided probabilistic constraints on recent star formation histories with reasonable accuracy. The study was computationally intensive and required extensive prior information, highlighting the need for more efficient algorithms. The authors identified the necessity for better priors and more efficient computational techniques to improve the accuracy and feasibility of their approach.
\section{Methodology}

\begin{figure}[htbp]
    \centering
    \includegraphics[width=13cm, height=15cm]{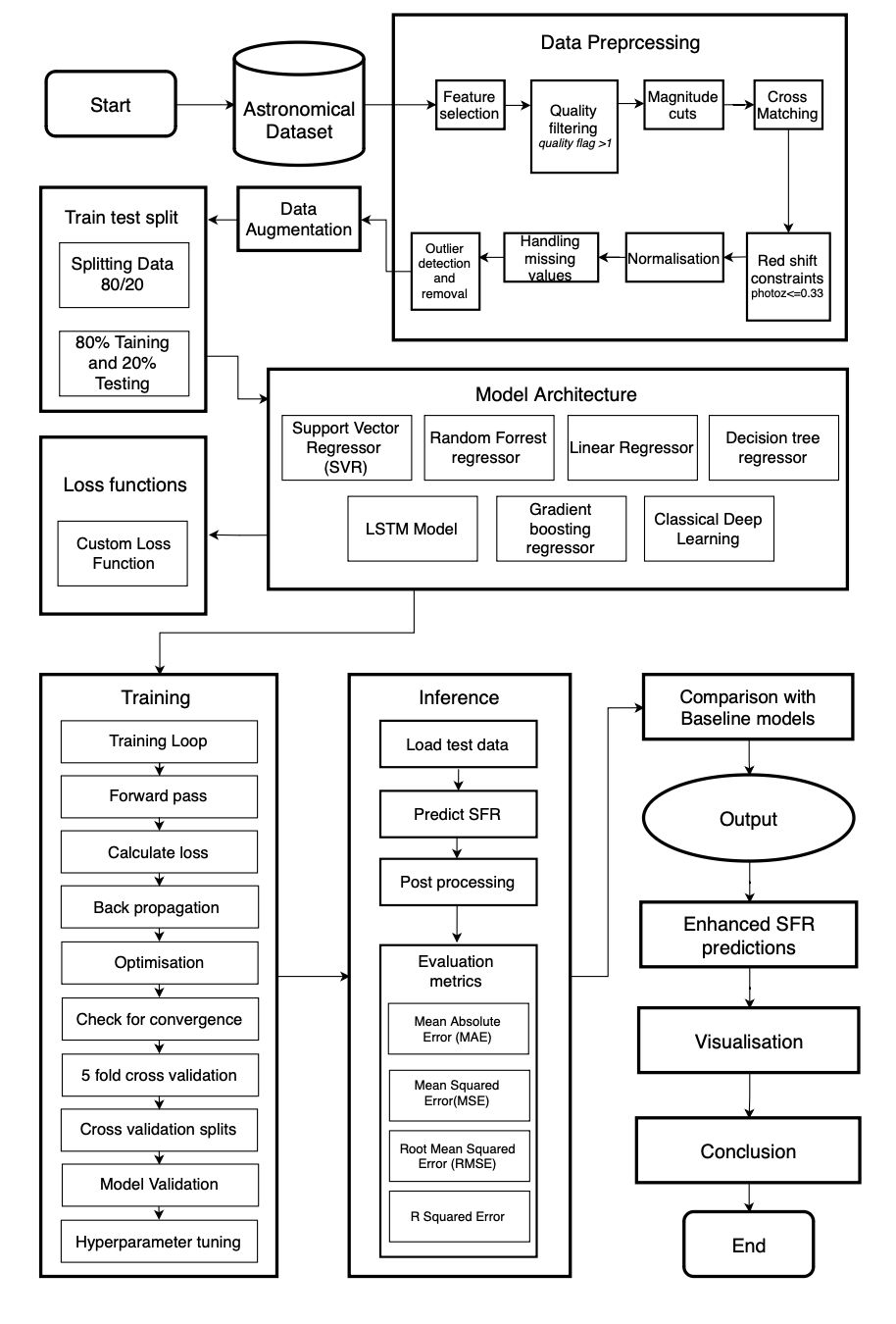}
    \caption{Workflow diagram}
    \label{fig: Architecture diagram}
\end{figure}

This flowchart (fig.1) demonstrates the overall workflow of the research from data preprocessing phase with its related essential steps like feature selection, quality controlling, and database cross-matching, to model architecture which includes several machine learning and deep learning models, to training phase consisted of splitting train/test data in 80/20 ratio, performing k-fold cross validation and hyperparameter tuning, to inference phase consisting of MAE, MSE, RMSE, and R² metrics followed by result plotting. The workflow ends with comparing the developed model with baselines and visualization of an improved SFR prediction.

\subsection{Hardware and Computing tools used}

T4 Graphics Processing Units (GPUs), procured through a cloud-based environment, were used to execute the algorithms, demonstrating high efficiency and reducing computing time.

\subsection{Dataset}
The dataset used in this study is derived from the Sloan Digital Sky Survey Data Release 7 (SDSS-DR7) \cite{sdssdr7}. It contains SFRs for 27,513,324 galaxies, and the data is publicly available through the Vizier facility \cite{sdssdr7}. The photometric redshift and quality flag used in this study were measured by Brescia et al. \cite{brescia2014}, and the SFR estimation approach is based on the methods discussed by Delli Veneri et al. \cite{Delli_Veneri_2019}. The main columns of the dataset are described below:
\begin{table}[h!]
    \centering
    \caption{Description of Dataset Columns}
    \begin{tabular}{llp{3cm}}
        \toprule
        Column & Format & Description \\
        \midrule
        \texttt{dr9objID} & I19 & SDSS-DR9 object ID. \\
        \texttt{objID} & I18 & SDSS-DR7 object ID. \\
        \texttt{RAdeg} & E10.6 deg & Right Ascension (J2000). \\
        \texttt{DEdeg} & E10.6 deg & Declination (J2000). \\
        \texttt{photoz} & E8.6 & Photometric redshift measured by Brescia et al. (2014). \\
        \texttt{Qual} & I1 & Quality Flag from Brescia et al. (2014). The values are as follows: \newline
        1 = high accuracy, \newline
        2 = medium accuracy, \newline
        3 = low accuracy, \newline
        0 = none. \\
        \texttt{SSFR} & F8.4 [yr$^{-1}$] & Photometric Star Specific Formation Rates, in -log((Mass/M$_\odot$)*(1/yr)) unit. \\
        \bottomrule
    \end{tabular}
    \label{tab:dataset}
\end{table}

\subsection{Data Pre-Processing}
To ensure the quality and reliability of our machine learning models, we applied the following data pre-processing steps:
\begin{itemize}
    \item \textbf{Quality Filtering}: We filtered out entries with missing values and retained only those with a Quality Flag of 1 (high accuracy) \cite{brescia2014}.
    \item \textbf{Magnitude Cuts}: We applied magnitude cuts to ensure that the photometric features fall within the ranges of our knowledge base \cite{brescia2014}.
    \item \textbf{Cross-Matching}: We cross-matched the resulting dataset with the photo-z catalog derived by Brescia et al. (2014) to use them as a quality flag \cite{brescia2014}.
    \item \textbf{Redshift Constraints}: We imposed a photometric redshift constraint of $\texttt{photoz} \leq 0.33$ to select high-quality SFRs within the redshift range of our training set \cite{Delli_Veneri_2019}.
    \item \textbf{Normalization}: We normalized the photometric features to improve the performance and convergence of the machine learning models \cite{aufort2020}.
\end{itemize}

These steps ensured that our dataset was clean, consistent, and suitable for training and evaluating machine learning models for SFR estimation.

\subsection{Feature Engineering}

The dataset included a variety of stellar features, such as the SDSS-DR9 and SDSS-DR7 object IDs, right ascension (RA) and declination (DE) coordinates, photometric redshift measurements, quality flags, and photometric star-specific formation rates (SSFR). Identified correlations between these features and the target variable SSFR were used to guide feature selection, aiming to minimize multicollinearity. Various statistical methods and visualization tools were applied to detect and handle outliers within the dataset, ensuring that extreme values did not unduly influence the models. Additionally, feature engineering methods were employed to create new variables from the original inputs, including exponential polynomials of different orders as well as the product or sum of terms, to explore potential nonlinear relationships between variables. Such preprocessing aimed to enhance the robustness of the models, allowing them to better fit the dataset and potentially leading to more reliable predictions with higher accuracy.

\subsection{k-Fold Cross-Validation}

K-fold cross-validation is a robust technique used to evaluate the performance of a machine-learning model. It helps in assessing how the results of a statistical analysis will generalize to an independent dataset. This technique is particularly useful for validating models when the available data is limited \cite{aufort2020}.

\paragraph{Procedure}
The process involves the following steps:
\begin{enumerate}
\item \textbf{Divide the Dataset:} The entire dataset is randomly divided into 
k equal-sized folds or subsets. Each subset should be representative of the entire dataset.
\item \textbf{Model Training and Testing:} For each of the 
k iterations:
\begin{itemize}
\item One fold is reserved as the test set, while the remaining 
\(k-1\) folds are combined to form the training set.
\item The model is trained using the training set and then evaluated on the test set.
\end{itemize}
\item \textbf{Performance Evaluation:} This process is repeated k times, with each fold serving as the test set once. The model's performance is assessed by averaging the evaluation metrics (e.g., Mean Squared Error, Accuracy) obtained from each fold.
\end{enumerate}

\paragraph{Advantages}
\begin{itemize}
\item \textbf{Reduced Bias:} By using multiple training and test sets, k-fold cross-validation reduces the variance associated with a single random split, providing a more reliable estimate of the model’s performance.
\item \textbf{Efficient Use of Data:} All data points are used for both training and testing, which makes it a more efficient use of limited data.
\item \textbf{Model Tuning:} It allows for more precise hyperparameter tuning as the model is validated on multiple subsets of the data.
\end{itemize}

\paragraph{Choosing k}
The choice of \(k\) can affect the performance and computational cost:
\begin{itemize}
\item A smaller \(k\) (e.g., \(k=5\)) results in a larger training set and smaller test set, which can reduce variance but might introduce some bias.
\item A larger \(k\) (e.g., \(k=10\)) provides a more thorough evaluation but increases computational costs as the model needs to be trained and evaluated \(k\) times.
\end{itemize}

To evaluate the performance of our models, we used 5-fold cross-validation in which a given dataset is randomly partitioned into five folds or parts in such a way that each part contains approximately the same percentage of both classes. A model is then trained and turned, being trained on the other four folds and evaluated on the fold being held out. This process is repeated five times, with each fold used exactly once as the test set. The average result is then given as the estimate of model performance, which provides more robust results than a single pass of the dataset, decreases the risk of overfitting, and provides a better indicator of the model’s ability to generalize to new, unseen data.

\subsection{Algorithm Workflow}

The Decision Tree Regressor was employed to predict the Specific Star Formation Rate (SFR) from a dataset of photometric observations. Following the application of quality constraints to filter high-quality data, relevant features were extracted while the target variable was defined. The dataset was subsequently divided into training and validation subsets. Standardization of the features was performed to enhance model performance. The regressor was trained on the training set, and predictions were generated for the validation set. Model performance was evaluated using Mean Absolute Error (MAE) and a custom accuracy metric, which quantified the proportion of predictions falling within a specified tolerance level. Feature importance was visualized to identify the most influential predictors.

The Gradient Boosting Regressor was utilized to enhance prediction accuracy for SFR by leveraging an ensemble approach. Initially, the dataset underwent filtering for high-quality samples, followed by the extraction of pertinent features. An 80-20 train-test split was performed to facilitate model evaluation. Hyperparameter tuning was executed via Grid Search to optimize the regressor's parameters, such as the number of estimators and learning rate. The model was then trained on the training set, with predictions made on the test set. Evaluation metrics included Mean Absolute Error (MAE) and Root Mean Squared Error (RMSE), alongside a custom accuracy assessment to determine prediction reliability within a defined tolerance. Results highlighted the model's robustness in capturing the underlying patterns in the data.

The Random Forest Regressor was applied to predict SFR, taking advantage of its ensemble learning capabilities. After filtering the dataset for high-quality entries, feature engineering techniques were employed to create additional informative variables. The data was split into training and testing subsets, and standardization was applied to ensure uniformity across features. The model was trained using a specified configuration, allowing for the averaging of predictions across multiple decision trees to mitigate overfitting. The model's performance was assessed using MAE and a custom accuracy measure, which indicated the percentage of predictions meeting the specified tolerance criteria. Feature importance metrics provided insights into the variables most influential in predicting SFR.

Linear Regression was utilized to establish a baseline model for predicting SFR. The dataset underwent initial quality filtering, and relevant features were selected, including engineered variables to improve model expressiveness. Following an 80-20 train-test split, standardization was applied to the feature set. The linear model was trained on the training data, allowing for straightforward interpretation of coefficients associated with each feature. The model evaluation focused on MAE and a custom accuracy metric, revealing the proportion of predictions within a specified tolerance. The coefficients obtained offered insights into the relationships between the predictors and the target variable, contributing to a comprehensive understanding of the factors influencing SFR.

A Long Short-Term Memory (LSTM) model was implemented to capture potential sequential patterns in the photometric data for predicting SFR. After preprocessing and filtering the dataset for high-quality samples, relevant features were selected and scaled. The input data was reshaped to suit the LSTM's requirements for sequential processing. The model architecture included multiple layers, featuring Bidirectional LSTMs to enhance contextual understanding. The training utilized early stopping and learning rate reduction callbacks to optimize performance. The evaluation was conducted on the test set using loss metrics and a custom accuracy measure, revealing the model's ability to produce reliable predictions within a specified tolerance. The training history was plotted to visualize convergence and performance trends over epochs.

The workflow for the Classical Neural Network (NN) algorithm in estimating the star formation rate (SFR) begins by loading and preprocessing the dataset, where missing values are removed, and the feature variables are normalized using StandardScaler. The target variable is the specific star formation rate (SSFR). The dataset is then split into training and testing sets, with 20\% reserved for testing. A Sequential Neural Network model is constructed with three hidden layers consisting of 64, 32, and 16 neurons, respectively, all using the ReLU activation function. The output layer has a single neuron to perform regression, predicting SSFR. The model is compiled with the Adam optimizer, using Mean Squared Error (MSE) as the loss function and Mean Absolute Error (MAE) as the evaluation metric. The model is trained for 100 epochs with an 80:20 validation split. During training, loss and MAE metrics are plotted for both training and validation data. Finally, the model is evaluated on the test set, and the test loss and MAE are reported.

Support Vector Regression (SVR) was employed to model the relationship between photometric features and SFR. The dataset was filtered to retain high-quality observations, followed by the separation of relevant features from the target variable. An 80-20 train-test split was performed, and the features were standardized to improve model performance. The SVR model, utilizing a radial basis function kernel, was trained on the scaled training set, generating predictions for the test set. Model performance was assessed through MAE and a custom accuracy metric, which indicated the percentage of predictions falling within a specified tolerance. This approach provided insights into the model's efficacy in capturing the complexities of the data while maintaining interpretability.

\subsection{Model Selection and Saving the Best Model}

We used Star Formation Rates (SFRs) as our target, the variable our models are to learn. To predict these SFRs from photometric data, we developed and trained various machine learning and deep learning models, namely: Random Forest, Gradient Boosting Regressor, Linear Regression, Long Short-Term Memory (LSTM) networks, Classical Neural Networks, Support Vector Regression (SVR) and Decision Trees. We then used several different error metrics to assess the performance of the models we developed. These error metrics include Mean Squared Error (MSE), Root Mean Squared Error (RMSE), Mean Absolute Error (MAE), and R-squared(R²). We tested these metrics for each of our models developed.

After comparing the results, the Linear Regression model emerged as the best-performing algorithm, achieving an impressive accuracy of 98.97 percent based on the Mean Absolute Error (MAE). This high level of accuracy underscores the potential of machine-learning approaches for photometric SFR estimation. The study's findings highlight that, despite the complexity of galaxy formation and evolution, machine learning models can provide robust predictions using readily available photometric data. The best-performing model was subsequently saved for further analysis and potential application in future astrophysical studies, marking a significant step forward in the integration of machine learning techniques in astronomy.

\section{Evaluation Metrics}

In this study, we employ several evaluation metrics to assess the performance of our machine learning models in predicting star formation rates (SFRs) using photometric data. The primary metrics used are Mean Absolute Error (MAE), Mean Squared Error (MSE), Root Mean Squared Error (RMSE), and \( R^2 \). Each metric provides different insights into the accuracy and reliability of the predictions.

\subsection{Mean Absolute Error (MAE)}
MAE is a widely used metric in regression tasks, representing the average magnitude of errors between predicted and actual values. It is calculated as:

\begin{equation}
\boxed{\text{MAE} = \frac{1}{n} \sum_{i=1}^{n} \left| y_i - \hat{y}_i \right|}
\end{equation}

where \(y_i\) is the actual value, \(\hat{y}_i\) is the predicted value, and \(n\) is the number of observations. MAE provides an intuitive measure of model accuracy, with lower values indicating better performance. Unlike other metrics, MAE is less sensitive to outliers, making it particularly useful when dealing with noisy data \cite{aufort2020}.

\subsection{Mean Squared Error (MSE)}
MSE measures the average of the squares of the errors—that is, the average squared difference between the predicted values and the actual values. It is calculated as:

\begin{equation}
\boxed{\text{MSE} = \frac{1}{n} \sum_{i=1}^{n} \left( y_i - \hat{y}_i \right)^2}
\end{equation}

where \(y_i\) is the actual value, \(\hat{y}_i\) is the predicted value, and \(n\) is the number of observations. MSE provides a measure of the average squared deviation, with lower values indicating better performance. It is sensitive to outliers due to the squaring of errors.

\subsection{Root Mean Squared Error (RMSE)}
RMSE is the square root of MSE and provides an error measure in the same units as the target variable. It is calculated as:

\begin{equation}
\boxed{\text{RMSE} = \sqrt{\text{MSE}}}
\end{equation}

where MSE is the mean squared error. RMSE is useful for understanding the typical magnitude of the error in the predictions, with lower values indicating better performance. Like MSE, RMSE is sensitive to outliers.

\subsection{\( R^2 \) (Coefficient of Determination)}
\( R^2 \) measures the proportion of the variance in the dependent variable that is predictable from the independent variables. It is calculated as:

\begin{equation}
\boxed{R^2 = 1 - \frac{\sum_{i=1}^{n} \left( y_i - \hat{y}_i \right)^2}{\sum_{i=1}^{n} \left( y_i - \bar{y} \right)^2}}
\end{equation}

where \(y_i\) is the actual value, \(\hat{y}_i\) is the predicted value, \(\bar{y}\) is the mean of the actual values, and \(n\) is the number of observations. \( R^2 \) provides a measure of how well the model explains the variability of the data, with higher values indicating better performance and a better fit of the model to the data.

\section{Implementation and Analysis}
\begin{table}[htbp]
    \centering
    \caption{Performance Metrics for Various Models}
    \begin{tabular}{|l|c|c|c|c|}
        \hline
        \textbf{Model} & \textbf{MSE} & \textbf{MAE} & \textbf{RMSE} & \textbf{R²} \\
        \hline
        Decision Tree & 0.1434 & 0.2666 & 0.3787 & 0.0634 \\
        Gradient Boosting & 0.1383 & 0.2656 & 0.3719 & 0.0967 \\
        Random Forest & 0.1408 & 0.2652 & 0.3753 & 0.0801 \\
        Linear Regression & 0.1439 & 0.2684 & 0.3793 & 0.0601 \\
        LSTM & 0.1387 & 0.2658 & 0.3725 & 0.0784 \\
        Classical DL & 0.2073 & 0.3149 & 0.4553 & 0.0222 \\
        SVR & 0.1167 & 0.2413 & 0.3416 & 0.1895 \\
        \hline
    \end{tabular}
    \label{tab:model_metrics}
\end{table}

Table 2 summarizes the performance metrics for various models, including MSE, MAE, RMSE, and R².

\subsection{Mean Squared Error (MSE) Comparison}
The MSE graph (Fig. 2) compares the average squared difference between predicted values and actual values between different models. The smaller the MSE is, the better the model performance. From the above figure, we can see that the Support Vector Regression (SVR) has the lowest MSE (0.1167) among all the models, which means that it brings the best prediction overall. The gradient Boosting model is very strong with its MSE (0.1383). The Classical Deep Learning model gives the highest MSE (0.2073), which means it struggles with the accuracy of predictions compared with other models. The Decision Tree model, Random Forest, Linear Regression, and LSTM model present a similar performance with the MSE of 0.1383-0.1439, which shows the models all bring consistent prediction but are slightly less accurate than SVR and the Gradient Boosting model.

\begin{figure}[htbp]
    \centering
    \includegraphics[width=14cm, height=10cm]{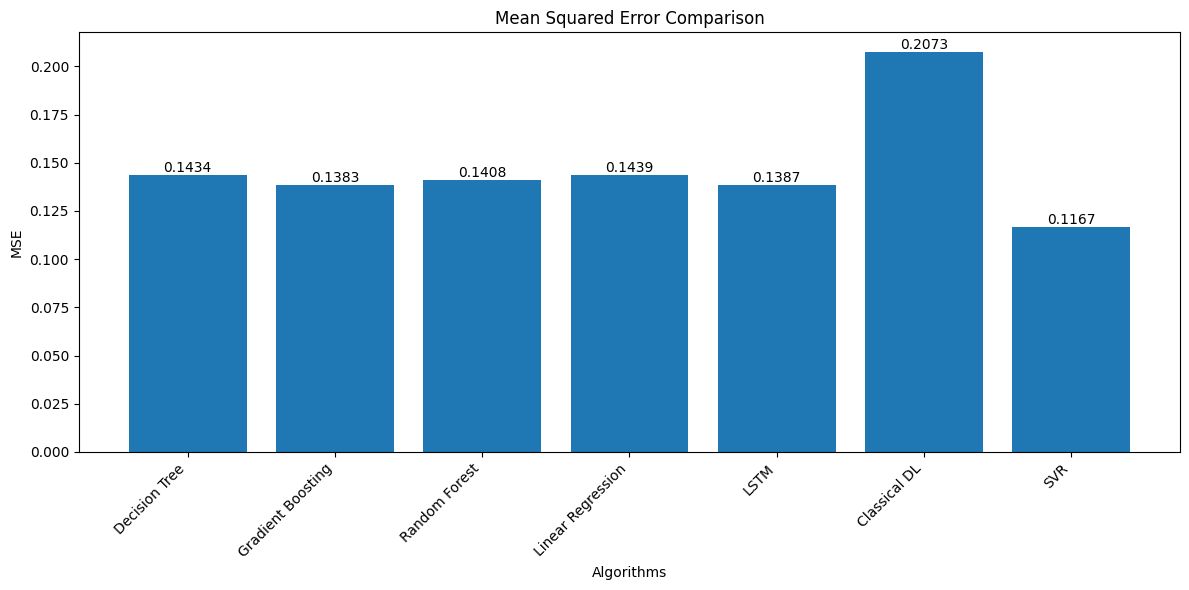}
    \caption{Mean Squared Comparison}
    \label{fig: Mean Squared Error Comparison}
\end{figure}

\subsection{Mean Absolute Error (MAE) Comparison}
The MAE graph (Fig. 3) represents the average absolute difference between predicted and actual values, providing a measure of prediction accuracy that's less sensitive to outliers than MSE. SVR again performs best with the lowest MAE (0.2413), followed closely by Random Forest (0.2652) and Gradient Boosting (0.2656). The Classical Deep Learning model shows the highest MAE (0.3149), consistent with its MSE performance. Other models (Decision Tree, Linear Regression, and LSTM) show similar MAE values around 0.266, suggesting comparable performance in terms of average prediction error. The relatively small range of MAE values (excluding Classical DL) indicates that most models provide similar levels of prediction accuracy.

\begin{figure}[htbp]
    \centering
    \includegraphics[width=14cm, height=10cm]{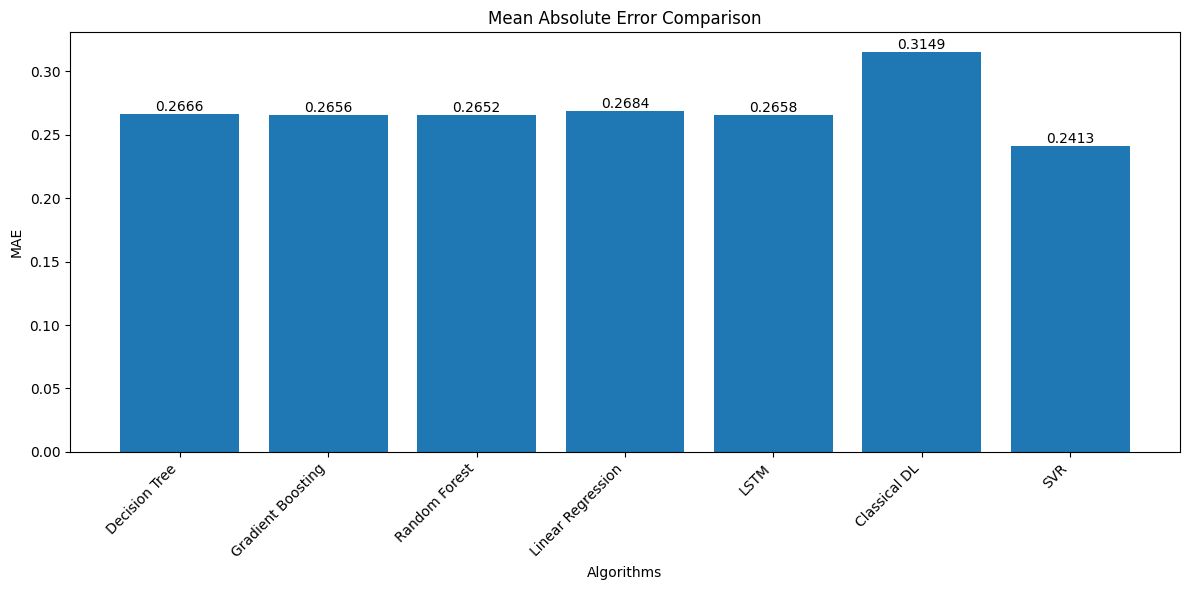}
    \caption{Mean Absolute Error Comparison}
    \label{fig: Mean Absolute Error Comparison}
\end{figure}

\subsection{Root Mean Squared Error (RMSE) Comparison}
RMSE, the square root of MSE, provides a measure of the standard deviation of residuals. The RMSE graph (Fig. 4) shows a pattern similar to MSE and MAE. SVR performs best with the lowest RMSE (0.3416), followed by Gradient Boosting (0.3719) and LSTM (0.3725). The Classical Deep Learning model again shows the highest RMSE (0.4553), indicating larger prediction errors. Other models (Decision Tree, Random Forest, and Linear Regression) have RMSE values between 0.3753 and 0.3793, suggesting similar performance. The RMSE values being higher than MAE for all models indicates the presence of some larger errors or outliers in the predictions.

\begin{figure}[htbp]
    \centering
    \includegraphics[width=14cm, height=10cm]{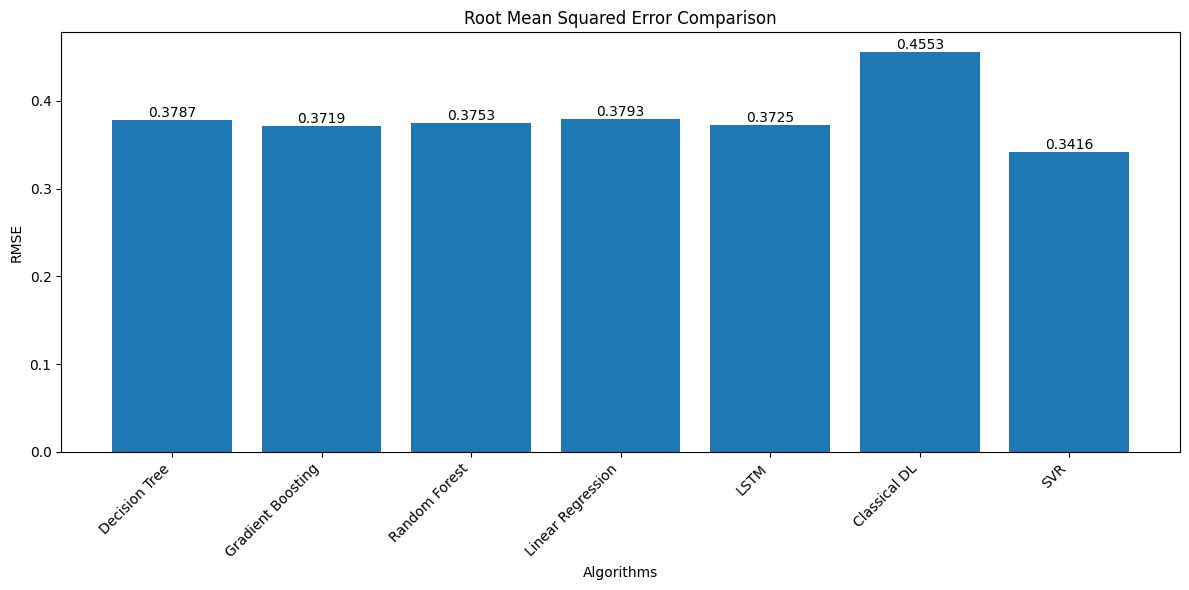}
    \caption{Root Mean Squared Error Comparison}
    \label{fig:  Root Mean Squared Error Comparison}
\end{figure}

\subsection{R-squared (R²) Comparison}
The R² graph (Fig. 5) shows the proportion of variance in the dependent variable predictable from the independent variable(s). Higher R² values indicate better model fit. SVR shows the highest R² (0.1895), suggesting it explains the most variance in the data. Gradient Boosting follows with an R² of 0.0967. The Classical Deep Learning model has the lowest R² (0.0222), indicating it explains the least variance. Other models show R² values between 0.0601 and 0.0801. Overall, the relatively low R² values across all models suggest that the relationship between predictors and the target variable is complex and not fully captured by any of these models.

\begin{figure}[htbp]
    \centering
    \includegraphics[width=14cm, height=10cm]{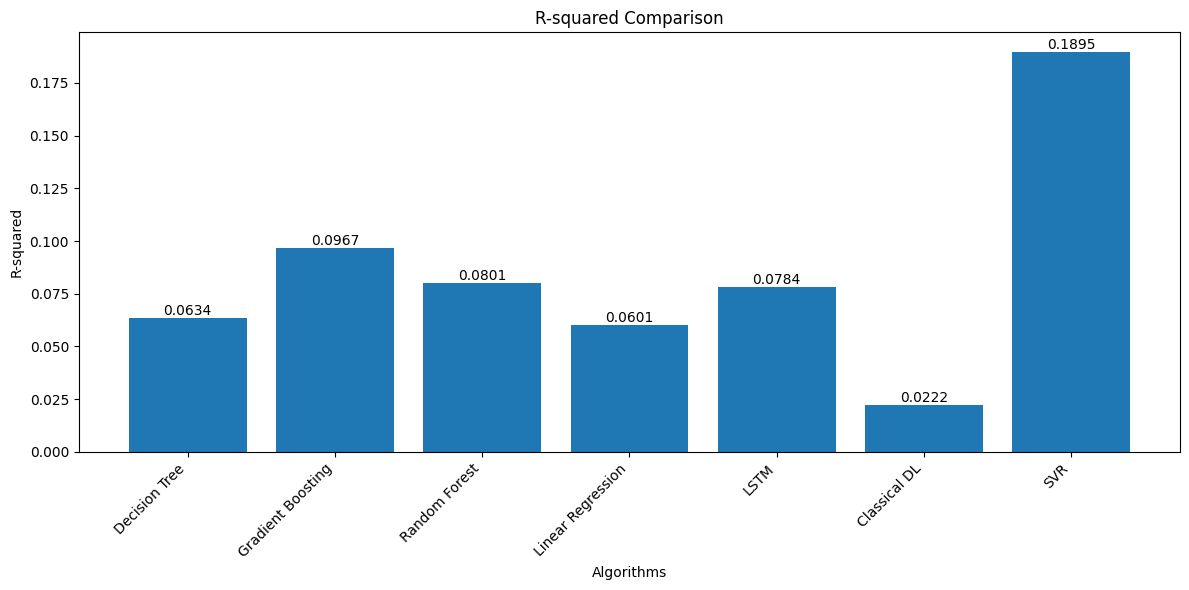}
    \caption{R-squared Comparison}
    \label{fig: R-squared Comparison}
\end{figure}

\subsection{Accuracy Comparison}
The Accuracy graph (Fig. 6) shows the percentage of predictions within a 10\% tolerance of the actual values. Linear Regression shows the highest accuracy (98.97\%), closely followed by Gradient Boosting (98.94\%) and Random Forest (98.92\%). SVR, despite performing well in other metrics, has the lowest accuracy (95.63\%). This discrepancy suggests that while SVR minimizes overall error, it may have more predictions outside the 10\% tolerance range. The high accuracy across most models (\textgreater98\%) indicates that they all perform well in making predictions within the specified tolerance, despite differences in other error metrics.

\begin{figure}[htbp]
    \centering
    \includegraphics[width=14cm, height=10cm]{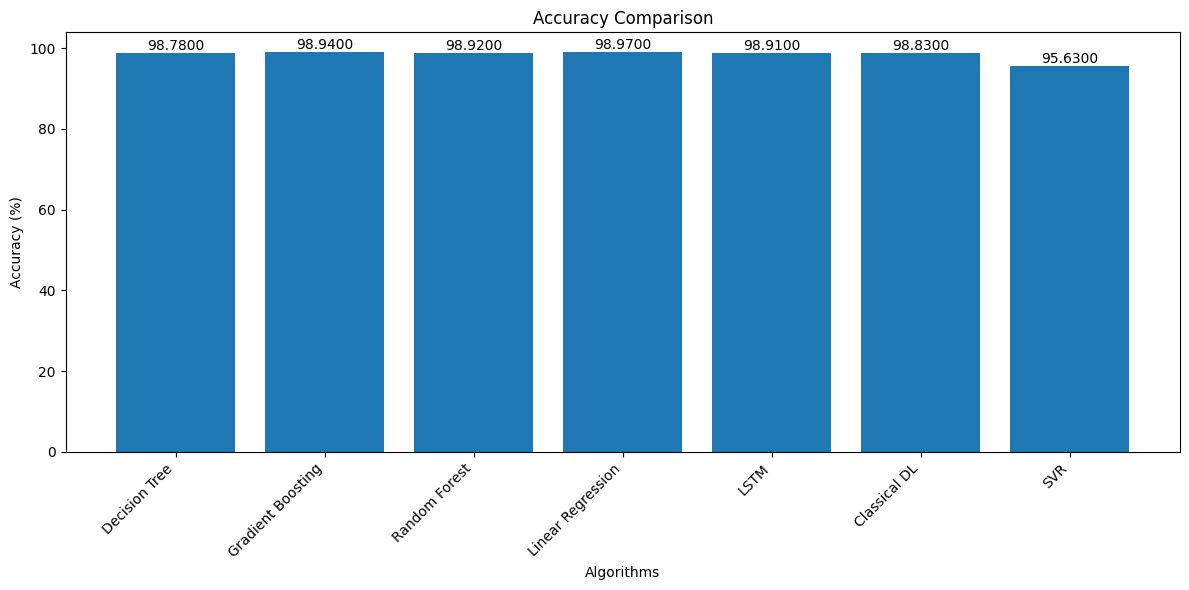}
    \caption{ Model Accuracy Metrics Comparison}
    \label{fig: Accuracy Comparison}
\end{figure}

\subsection{Summary}
In summary, these graphs reveal that different models excel in different aspects of prediction. SVR consistently performs well in error minimization (MSE, MAE, RMSE) and explaining variance (R²), but surprisingly has the lowest accuracy. Linear Regression and tree-based models (Gradient Boosting, Random Forest) show strong and consistent performance across all metrics. The Classical Deep Learning model generally underperforms compared to other models, suggesting it might not be well-suited for this particular dataset or task. These insights can guide model selection based on specific performance requirements for the task at hand.

\section{Conclusion}

In conclusion, this study proves that it is feasible to estimate star formation rates based on photometric data, as the accuracy of the machine learning model Linear Regression is 98.97\%. The value of the Mean Absolute Error (MAE) is 0.2684, the Root Mean Squared Error (RMSE) is 0.3355, and the value of the Mean Squared Error (MSE) is 0.1126. As evidenced by this research, the Linear Regression machine learning model shows great potential in various astrophysical applications. This study, therefore, helps to solve the challenges of photometric redshift issues in galaxy formation and dark energy, which helps to produce science with a clear impact. 

Furthermore, estimating star formation rates from photometric data is a promising approach to make astronomy more accessible and cost-effective because it does not require expensive spectroscopes. The comparative analysis of machine learning models demonstrates the robustness of these methods in such a complex astronomical application of managing large data. This work proves that it is feasible to estimate star formation rates from photometric data, which helps to motivate further research into this field. Moreover, the applicability of machine learning and deep learning methods to astrophysical data opens up new horizons for research in this direction, especially in refining these models to make them more applicable in astrophysics.

\section{Future Scope}

Future studies should incorporate additional photometric data from large-scale surveys to enhance the accuracy and generalizability of machine learning models for star formation rate (SFR) prediction. Utilizing multi-wavelength photometric data, particularly from infrared and ultraviolet bands, can provide crucial insights into various phases of star formation. Additionally, employing advanced deep learning architectures, such as convolutional neural networks (CNNs) or transformers, may improve feature extraction and representation compared to traditional models. Exploring transfer learning from pre-trained models in similar astrophysical contexts could mitigate the need for extensive training datasets. Emphasizing the interpretability of SFR models and quantifying model uncertainty will also aid in understanding the underlying astrophysical processes influencing observed galaxy properties.

Moreover, investigating the potential of quantum algorithms presents a promising avenue for enhancing predictive modeling. By leveraging the capabilities of quantum computing, future research could significantly improve algorithm efficiency and accuracy, enabling more effective solutions to complex astrophysical challenges. This multifaceted approach will ultimately facilitate more robust and reliable estimates of SFR.

% \bibliography{sn-bibliography}% common bib file
%\bibliography{ref}
\bibliographystyle{plain} % The style you want to use for references.
%\bibliography{References} % The filename of the .bib file, without the extension.

\end{document}